\documentstyle[11pt,a4,psfig]{article}
\begin{document}

\noindent
{\large\bf Interpretation of a low-lying excited state 
of the reaction center of Rb.sphaeroides as a double triplet}

\vskip 1cm

\noindent
{\bf P.O.J.Scherer and Sighart F. Fischer \hfill\\
Theoretical  Physics T38 \hfill\\
Technische Universit\"at M\"unchen \hfill\\
D-85748 Garching} \hfill\\

\subsection*{
Abstract:}
The recently observed transient absorption  of the lowest excited state
of the special pair $P^*$ at $2710 cm^{-1}$  [1]
 is assigned  as a singlet which 
arises from the coupling of the two lowest triplets from the two dimer 
halves. INDO calculations are used to predict its intensity. The analogy 
of the coupling mechanism to the trip-doublet spectrum from $P^+$ is shown
 and the influence of the double triplet on the Stark effect of $P^*$ is investigated.

\subsection*{
Introduction}

In recent femtosecond infrared spectroscopy of the reaction center 
excited states of Rb.sphaeroides up to $10000 cm^{-1}$ an interesting 
spectrum has been observed. A relatively sharp band at $2710 cm^{-1}$ is 
followed by a very broad around $5300 cm^{-1}$ [1 ]. These bands have been 
interpreted as internal charge transfer states, where the extra broadening 
of the upper is seen as a result of mixing to higher excited states of 
the dimer. This interpretation can not explain that the higher state has 
more intensity, also the absence of the lower in the ground state spectrum
 and in the Stark effect spectrum would be in conflict with this model, 
since the CT state should mix appreciably with $P^*$. From our INDO calculations 
which are based on the X-ray structure of rps.viridis we have predicted 
earlier [ 2 ] the presence of a low lying excited state denoted $P^{**}$ which
 is a singlet component of a double excited state with a triplet on each
 of the two dimer halves. We obtained this state too low in energy as we 
know from the predicted energy positions of the lowest triplet. However,
 if we correct for this shortcoming of the INDO calculation, we predict 
this state to be at the position of the newly observed band. For the 
symmetric dimer it is an even state and therefore does not couple with $P^*$. 
There exists, however, an interesting one-electron coupling mechanism to the 
internal charge transfer states. This coupling is of the same nature as the 
one which provides intensity to the trip-doublet observed in the $P^+$ spectrum. 
In this paper we want to show that the $P^* \rightarrow P^{**}$ transition obtains its 
intensity from the CT states. Moreover the presence of the $P^{**}$ state helps 
to understand the unusual Stark effect spectrum of $P^*$ and provides insight 
into the internal asymmetry of the dimer. We start with the analysis of the
 spectrum of $P^+$, then we present an extended exciton model for the dimer 
which includes the CT and the double-triplet state. Analytical expressions
 are derived for the infrared transitions and the Stark effect. The relevant
 couplings are discussed in the framework of INDO calculations.

\subsection*{
Simulation of the $P^+$ spectrum}
The spectrum of $P^+$ [1,3] shows a very low energy transition at 
$\approx 2500 cm^{-1}$ with  $80 Debye^2$ intensity and another transition of 
similar strength at  $ \approx   7500 cm^{-1}$ with an additional shoulder at 
higher transition energy.  INDO calculations by us [2] as well as model 
studies by Breton et al [3] and by Parson et al [4 ] related the lowest 
transition to the hole transfer. The higher transition shows more complex 
configuration interaction [2]. Recent investigations by Reimers et al  [5 ]
attributed also the low energy transition to the hole transfer between the 
two dimer halves and the higher band to a trip-doublet state. A quantitative 
understanding is still missing since the internal asymmetries of the dimer
 wich  are difficult to assess, influence the relative intensities in a sensitive way.

Here we want to discuss the experimental spectrum in the framework of a simplified 
four state model which incorporates the two cationic states $L^+$ and
$M^+$   and the two
 lowest excitations for each of them which are the triplet-coupled doublets
 $L^+M^3$ and $M^+L^3$. A more detailed analysis including higher excited configurations 
explicitly will be given in a separate paper .

\begin{equation} H=\left(
\begin{array}{cccc}
E_1 - d_1  & U_1  &  0  &  U_{12}  \\
U_1 &  E_1 + d_1  & U_{21} & 0     \\
0  &  U_{21} &  E_2 - d_2  &  U_2  \\
U_{12}  &  0  &  U_2  & E_2 + d_2  \\
\end{array}
\right) \end{equation}

Within this basis the positive charge is fully localized on $P_L$ for the first and
 third state and on $P_M$ for the other two. The delocalization of the charge is the 
result of the intermolecular resonance interaction $U_1$ ,$U_2$ competing with  the 
small energy differences $2d_{1,2}$ of the symmetry related states. Such asymmetries 
arise from asymmetric interaction with the surrounding, e.g. hydrogen bonding 
to protein residues or from asymmetries of the two chromophores.  We investigated
 especially the influence of the orientation of the two acetyl groups. A structure
 model where the oxygen and the methyl group of  $P_M$ are exchanged is also 
consistent with the X-ray structure and leads to a large splitting of the intradimer
 charge transfer states which is needed to explain the strong dipole change of $P^*$ 
seen in the Stark spectrum. The two couples of states denoted by the indices 1 and 2 
respectively interact via the matrixelements $U_{12}$ and $U_{21}$.  A detailed analysis of 
the INDO results shows that the coupling $U_1$ is mainly given by the negative 
resonance interaction $-\beta_{LM}$ involving the molecular orbitals of two different 
charged states $P_L^+$ and $P_M^+$.  For our analysis we want to interpret couplings 
for $P^+$ and $P^*$ within one set of molecular orbitals. We use localized SCF orbitals 
of the neutral dimer which are obtained switching off all resonance interactions
 between the two dimer halves during the SCF step. Electron correlation effects 
are important and the expression for $U_1$ becomes more complicated as   configuration
 mixing   takes place. We find that $U_1$ is underestimated by about a factor of 2
 if the resonance integral $\beta_{LM}$  of the SCF orbitals of P is used. For the coupling
 of the two trip-doublet states the reorganization effect is of little importance.
 The couplings of the lowest cationic states to the trip-doublets involve the
 transfer of an electron from the HOMO on one side to the LUMO on the other side.
Here the reorganization effect leads to a significant increase of $U_{12}$ and $U_{21}$. 
These rather strong couplings induce the intensity of the trip-doublets. The results 
are summarized in table 1

\noindent
For a simplified analytical treatment we first diagonalize the two pairs of states with 
their asymmetries which we want to determine empirically. The couplings $U_{12}$ and $U_{21}$ are
 treated as a perturbation. We use the linear combinations of states

\begin{equation} \tilde H = S^{-1} H S = \left(
\begin{array}{cccc}
\cos(\phi_1) & \sin(\phi_1) & & \\
-\sin(\phi_1) & \cos(\phi_1) & & \\
 & &	\cos(\phi_2) & \sin(\phi_2)\\
 & &    -\sin(\phi_2) & \cos(\phi_2) \\
\end{array}
\right) \end{equation}

where the angles $\phi_{1,2}$  measure the degree of localization or delocalization and are given by

\begin{equation}\sin(\phi_j) = - sign(U_j) 
{\sqrt 2 \over 2}
 \sqrt{  {\sqrt{d_j^2 + U_j^2}-d_j}   \over{\sqrt{d_j^2+U_j^2}}  } 
\qquad  \cos(\phi_j)=\sqrt{1-\sin(\phi_j)^2} \end{equation}

Specifically one obtains $\phi_j = 0 (\pm {\pi \over 2})$ if the lower state of a pair is localized on $P_{L(M)}$  and $\phi_j = \pm {\pi \over 4} $  if 
 both states of the pair are fully delocalized. Within this model we can determine the 
main parameters empirically and compare them with our calculated ones. From the experimentally
 observed charge distribution of $ \approx 65 \%$ on $P_L$ we get
$ \sin^2(\phi_1)\approx 0.35 $.

  The intensity of the lowest transition
 is estimated as
$({p \over 2} \sin(2 \phi_1))^2 \approx 0.23 p^2$. 
 If we take a value of p=38 Debye corresponding to the full center to 
center distance the calculated intensity of $330 Debye^2$ is  too large by a factor of 
about 3 to 4. The quantum calculations, however, reduce  p for the cationic states by 
the rearrangement of the electron system to a value of about 20 Debye and the calculated
 intensity ($90 Debye^2$) is close to the experimental data of $80 Debye^2$ for viridis. From 
the experimental transition energies we find approximately
$2 \sqrt{d_1^2 + U_1^2} \approx 2700 cm^{-1}$.
Together with
${1 \over 2}\left( {1 - {d_1 \over \sqrt{d_1^2 + U_1^2}}}\right) = \sin^2(\phi_1) \approx 0.35$
 we estimate the
 coupling and  energy difference as $U_1 = 1300 cm^{-1}$ and
 $2d_1=800 cm^{-1}$. 
The value of $U_1$ is
 larger than the calculated $800cm^{-1}$. This discrepancy could be partly due to a geometry
 relaxation in the cationic state but  also to the parameterization of the resonance 
integrals which was optimized with respect to transition energies. In our simulation
 (fig.1a) we accounted for this shortcoming and used the larger value of 
$1300 cm^{-1}$.
If we assume that the intensity of the trip-doublets is mainly borrowed from the hole transfer 
transition  the total intensity of the two trip-doublets is given by 
$p  U_{12}^2 /(E_2-E_1)^2  \approx  25 D^2$
 for the calculated coupling of $1600 cm^{-1}$ and an energy difference of
$ 7000 cm^{-1}$. 
It depends sensitively on the value of the resonance coupling $U_{12}$.
 An increase of the coupling 
by a factor of 2 already gives more intensity for the trip-doublets than for the hole transfer band.  The distribution of intensity over the two trip-doublet transitions depends on their degree of localization. In case of full localization of the lower trip-doublet on $P_M$ the
 intensity ratio  of 2:1 is essentially  given by the localization of the lowest  doublets. 
 If on the other hand the two trip-doublets form a pair of completely delocalized states which
 might be a reasonable assumption as the resonance coupling $U_2$ is larger in magnitude
 than $U_1$, the lower of the trip-doublet transitions gains about
$ \sin^2(\phi_1+ {\pi \over 4}) = 97\%$ of the
 intensity and the higher transition becomes rather weak.
The calculated splitting of the two trip-doublets is for the symmetric dimer
 $2 U_2 \approx  2000 cm^{-1}$
 but may be enlarged due to energy differences of the triplet states on
 the L and M halves. In our INDO calculations we found a rather small asymmetry only if
 we used the structure with the rotated acetyl on the M-half. Experimentally it is not
 possible to assign this state clearly which supports the conjecture of strong delocalization
  and the resulting low intensity. This conclusion differs from the assumptions made
 by Reimers and Hush  [ 5 ].  If the trip-doublets borrow their intensity predominantly
 from the hole transfer band they will be essentially both polarized parallel to the vector p.
 This polarization is consistent with the experimental anisotropy of
 $0.298\pm0.028$ [1]. 
From the quantum calculations we find a value of 0.27. We like to point to the fact that 
the magnitude and direction of p are largely influenced by the reorganisation of the
 electron system. For a transition parallel to the direction connecting the two magnesium 
atoms the calculated anisotropy has a much smaller value of 0.21. On the basis of this 
assumption Wynne et al [1] concluded that the intensity of the trip-doublet is not 
provided by the hole transfer. Instead they assume that the trip-doublet has the polarization
 of the Qy transition of $P_M$.  The transition to the trip-doublet gains its intensity from
 the one-particle interaction between the two dimer halves as discussed by Reimers et al.
 A detailed analysis will be given in a forthcoming paper which shows that the transition 
dipole contains a contribution  from the hole transfer transition and a smaller one from 
the Qy transitions of the two halves, the latter depending on the degree of localisation 
of the $P^+$ states. 

\subsection*{
Interpretation of the $P^*$ spectrum}
The  electronic structure of the dimer in the Qy region is expected to show at least four 
states which are mixtures of two local excitations $P_L^*$ and $P_M^*$ and of two charge transfer 
states $L^+M^-$ and $M^+L^-$. The lowest excitation is mainly  ascribed to the antisymmetric
 combination $P_L^* - P_M^*$ with some admixture of the charge transfer states.  This excitation 
carries most of the Qy oscillator strength as the transition dipoles of the two dimer halves
 are nearly antiparallel. The  second excitonic component , the so called upper dimer band 
carries much less intensity and overlaps with the absorption bands of the accessory monomers. 
From the absence of a strong Stark effect in this region it was concluded [8] that the 
upper dimer band  couples only weakly to the CT states. The CT states are predicted to be 
in the region between the upper dimer band and the Qx transitions of the dimer. They cannot
 be observed in the absorption spectrum as they have only little intensity and  are 
inhomogeneously broadened by electrostatic interaction with the environment. In addition to
 these four basic Qy excitations we now incorporate an additional excitation in this 
frequency region which  can be visualized as two triplet excitations on the two halves which
 are coupled to a singlet double excitation of the dimer. It might be identified with the 
weak band at $15800 cm^{-1}$ denoted by a star in fig.4 of [ 1 ]. According to our INDO results 
this double-triplet state $P^{**}$ couples strongly to the charge transfer states and may be 
important for the interpretation of the Stark effect of the dimer absorption. The coupling 
involves the transfer of  an electron from an occupied orbital to an unoccupied orbital on 
the other dimer half and is therefore closely related to the coupling of the trip-doublets 
to the lowest cationic states. The energy of this state is underestimated by the INDO method 
for the applied parameters. As the lowest calculated triplet of the Bchl molecules comes out 
too low by about $1500 cm^{-1}$   and the calculated energy of $P^{**}$ is close to $P^*$ or even slightly
 below [2,9,10 ] we expect the correct position of $P^{**}$ somewhat less than $ 3000 cm^{-1}$   
 above $P^*$ which places it at the position of the lower and sharper $P^*$ absorption band of
 Wynne et al [1]. In the following we want to show that further support for such an assignment
 comes from the prediction of its intensity, its width and its Stark effect.
We consider  coupling of the following basic excitations: $P_L^*$,$P_M^*$, $L^+M^-$ ,
 $M^+L^-$  ,$P^{**}$ 
 described by the interaction matrix [ 2]

\begin{equation} H = \left(
\begin{array}{ccccc}
E(P_L^*) &  W  &  \beta_{L*M*}  &  -\beta_{LM} &  0  \\
W  & E(P_M^*)  &  -\beta_{ML} & \beta_{M*L*}  &  0  \\
\beta_{L*M*}  &  -\beta_{ML} & E(L^+ M^-) & 0 & -\sqrt{3 \over 2} \beta_{L*M} \\
-\beta_{LM} & \beta_{M*L*} & 0 & E(M^+ L^-) &   -\sqrt{3 \over 2} \beta_{M*L} \\
0 & 0 & -\sqrt{3 \over 2} \beta_{L*M} & - \sqrt{3 \over 2} \beta_{M*L} & E(P^{**})\\
\end{array}
\right) \end{equation}

The factor $\sqrt{3 \over 2}$  in front of the HOMO-LUMO cross coupling matrix elements results from the proper
 spin multiplicities. Contrary to $P^+$ the excited state $P^*$ is highly delocalized. So we start 
our analysis with a symmetric dimer with  
$ E(P_L^*)=E(P_M^*)=E(*)$ and $E(L^+M^- )=E(M^+L^-)=E(CT)$. 
The excitations can be classified  as symmetric or antisymmetric with respect to the $C_2$ symmetry
 operation.  The interaction matrix  reduces then to two blocks		

\renewcommand{\arraystretch}{2.0}

\begin{equation} 
\begin{array}{l}
\qquad H(-) =\left(
\begin{array}{cc}
E(P^*(-) & U_- \\
U_- & E(CT) \\
\end{array}
\right)\\

\noindent
\mbox{and}\\

\qquad H(+)=\left(
\begin{array}{ccc}
E(P^*(+) & U_+  & 0\\
U_+ & E(CT) &  U_d\\
0  &  U_d  &  E(P^{**}) \\
\end{array}
\right)\\

\noindent
 \mbox{with the new coupling matrixelements} 	

\quad U_- = \beta_{L*M*} + \beta_{LM} \quad U_+ = \beta_{L*M*}-\beta_{LM} 
\quad U_d = -\sqrt{6}\beta_{L*M} \\	

\noindent
\mbox{and the zero order excitonic energies }  

E(P^*(\pm))=E(*)\pm W. \\  	  
\end{array}
 \end{equation}

As the integrals $\beta_{LM} $ and $ \beta_{L*M*}$ involve  similar atomic  overlaps in the ring-I region they have 
the same sign. Therefore the coupling between excitonic and CT excitation is enhanced for
 the negative states whereas it is reduced for the positive ones , i.e. the upper dimer band 
will be largely decoupled from the two other positive excitations. In addition our INDO 
calculations gave a strong dynamic correlation effect for the upper dimer band largely reducing
 the coupling to the CT states. The coupling matrix elements  of the other states $U_-$ and $U_d$ show
 only minor changes as can be seen from table 2.

Since the coupling $U_+$ is so small we treat it as a perturbation. Diagonalizing the strongly coupled
 states we get as approximate eigenstates of the symmetric dimer:

\begin{equation}
\begin{array}{ll}
|P^*(-)> & =  {c_- \over \sqrt{2}} (P_L^* - P_M^*)
 + {s_- \over \sqrt{2}} (L^+M^- - M^+L^-) \\
|CT(-)> & = {c_- \over \sqrt{2}} (L^+M^- - M^+L^-)
  - {s_- \over \sqrt{2}}  (P_L^* - P_M^*)\\
|P^*(+)> & = {1 \over \sqrt{2}}  (P_L^* + P_M^*)\\
&+\left( 
\frac{c_+^2 U_+^2}{E(P^*(+))-E(CT(+))} 
+\frac{s_+^2U_+^2}{E(P^*(+)-E(P^{**})}
\right){1 \over \sqrt{2}}   (L^+M^- + M^+L^-)\\
& +  c_+s_+U_+
\left(
\frac{1}{E(P^*(+))-E(CT(+))}-\frac{1}{E(P^*(+)-E(P^{**})}
\right)
P^{**}\\
|P^{**}> & =  c_+ P^{**} +{s_+ \over \sqrt{2}}   (L^+M^- + M^+L^-)
 +  \frac{s_+U_+}{E(P^*(+))-E(P^{**})}{1 \over \sqrt{2}}(P_L^* + P_M^*)\\
|CT(+)> & =  {c_+ \over \sqrt{2}}   (L^+M^- + M^+L^-)-s_+P^{**}
 - \frac{c_+U_+}{E(P^*(+))-E(CT(+))}{1 \over \sqrt{2}}(P_L^* + P_M^*)\\
\end{array}
\end{equation}

where the mixing coefficients are given by ...		

\begin{equation}
\begin{array}{l}
s_- = -sign(U_-) \frac{\sqrt{2}}{2}
\sqrt{
\frac
{\sqrt{(E(CT)-E(P^*(-)))^2+ 4U_-^2}-(E(CT)-E(P^*(-))}
{\sqrt{(E(CT)-E(P^*(-)))^2+ 4U_-^2}} 
}\\
s_+ = -sign(\frac{U_d}{E(P^{**})-E(CT)}) \frac{\sqrt{2}}{2}
\sqrt{
\frac
{\sqrt{(E(CT(+))-E(P^{**}))^2+ 4U_d^2}-(E(CT(+))-E(P^{**}))}
{\sqrt{(E(CT(+))-E(P^{**})))^2+ 4U_d^2}} 
}\\

\end{array}
\end{equation}

Substituting the experimentally identified energies and the calculated couplings gives the 
ordering $s_-^2 < s_+^2 < c_+^2 < c_-^2$ . The intensities for transitions from the lowest excited 
state $P^* (-)$ are in this approximation

\begin{equation}
\begin{array}{ll}
\mu(P^*(-)\rightarrow P^*(+)) &= \frac{c_-}{2} (\Delta p_L - \Delta p_M)\\
&+ \left(
\frac {c_+^2U_+^2}{E(P^*(+))-E(CT(+))}
+ \frac{s_+^2 U_+^2}{E(P^*(+))-E(P^{**})}
\right) s_- p_{ct} \\

\mu(P^*(-)\rightarrow P^{**})&=s_+s_-p_{ct} + 
\frac{c_-s_+U_+}{E(P^*(+))-E(P^{**})} \frac{1}{2}(\Delta p_L - \Delta p_M)\\

\mu(P^*(-)\rightarrow CT(+))&=c_+s_-p_{ct} - 
\frac{c_-c_+U_+}{E(P^*(+))-E(CT(+))} \frac{1}{2}(\Delta p_L - \Delta p_M)\\

\mu(P^*(-)\rightarrow CT(-))&= 
 \frac{1}{2}c_-s_- (\Delta p_L + \Delta p_M)\\

\end{array}
\end{equation}

As the dipole changes $\Delta p$ of the isolated
 chromophores are small, the main contribution of the first three transitions comes from the
 large dipole moment of the charge transfer states $p_{CT}$ which from the INDO results has a
 magnitude of 32 Debye. So we predict for a nearly symmetric dimer  that only the transitions 
  to CT(+) and $P^{**}$ should carry significant intensities.

\subsection*{
Symmetry breaking interactions and Stark effect}

Breakage of the $C_2$ symmetry  causes a splitting of the two charge transfer states $L^+M^-$
and $M^+L^-$. Such perturbations easily arise from different orientation of the acetyl groups or
 from asymmetric interactions with the surrounding , including the local electric field. We model 
 the splitting of the two local CT states   by a perturbation operator of the form

\begin{equation}
\begin{array}{rcl}
H' & = & \frac{\delta}{2}
\left( 
|L^+ M^-><L^+ M^-| - |M^+ L^-><M^+ L^-| 
\right) \\
& = & \frac{\delta}{2}
(
 c_+s_- |P^*(-)><CT(+)| +c_+c_- |CT(-)><CT(+)|
 \\ 
& &  +  
s_-s_+ |P^*(-)><P^{**}| + c_-s_+ |CT(-)><P^{**}| + h.c.
 )
+ \dots
\end{array}
\end{equation}

Here $\delta$ is the perturbation  parameter which  admixes the states of different symmetry. 
The asymmetry does not effect in lowest order the intensities of the first three transitions in eq. 8. 
Only the transition to the fourth state CT(-) will borrow intensity from CT(+) as the two charge 
transfer states decouple into the localized states $L^+M^-$ and $M^+L^-$.
On the basis of these results we   interpret the $P^*$ spectrum in the following way: The transition 
to the upper band at $ \approx 1500 cm^{-1}$
 is difficult to assign since it is weak and overlaps with 
infrared active vibrations. The observed band at $2700 cm^{-1}$ represents the transition to the 
double-triplet state which is coupled strongly to the CT states. The  broader band at $5000 cm^{-1}$ 
contains transitions to the two CT states which are split and broadened by their interaction
 with the environment. This interpretation is also consistent with the different widths of the 
two bands. Only a small part of the width of the CT states is transferred. Figure 1b shows a 
simulation of the $P^*$ spectrum using the calculated couplings from table 2. The energies of the
 transitions had to be adjusted slightly to  fit the maxima of the experimental spectrum. Using
 the calculated couplings we have $s_-=0.37$ and$s_+=0.45$ . This gives a charge transfer contribution of $14 \%$ to 
the $P^*$(-) state and a total intensity of the $P^*$ spectrum of 
$ s_-^2 p_{ct}^2 \approx 140 Debye^2$
 which is already close to the 
experimental value of $ 200 Debye^ 2$ .  The  simulation with the parameters from table 3 agrees quite 
reasonably with the observed intensities  (fig.1).  Further support for this assignment comes 
from the Stark spectrum.

The energy of the lower
 dimer band depends on the asymmetry $\delta$. Using perturbation theory of  fourth order with respect
 to  $\delta$  and second order in $s_-$ and neglecting the zero order energy difference of the CT states
 we find

\begin{equation}
\begin{array}{l}

E(\delta)  =E(P^*(-))\\
 - {\left( {\frac{\delta}{2}} \right)}^2 s_-^2 
\left( {\frac{1}{E(CT)-E(P^*(-))} + 
 s_+^2 
\left( {\frac{1}{E(P^{**})-E(P^*(-))} 

-\frac{1}{E(CT)-E(P^*(-))}
}\right)
}\right) \\

-\left(\frac{\delta}{2}\right)^4 \frac{s_-^2}
{(E(CT)-E(P^*(-)))(E(P^{**})-E(P^*(-)))^2}
\left(1-c_+^2\frac{E(CT)-E(P^{**})}{E(CT)-E(P^*(-))}\right)^2 \dots 
\end{array}
\end{equation}

Even at zero electrical field asymmetric nuclear motions and static structural asymmetries 
contribute to the splitting of the CT states. Therefore we write $\delta$ as the sum of a field 
dependent and a constant term

\begin{equation}
\delta=\delta_0 +2 p_{ct}F 
\end{equation}

Expanding the energy as a function of the electric field we then have

\begin{equation}
\begin{array}{l}

E(F,\delta_0)=E(0,\delta_0) + p(\delta_0) F+ \alpha(\delta_0) F^2 + \dots \\

p(\delta_0)=-p_{ct} \delta_0 s_-^2 
\left(\frac{1}{E(CT)-E(P^*(-))}+s_+^2 \left(
\frac{1}{E(P^{**}-E(P^*(-))}-\frac{1}{E(CT)-E(P^*(-))}
\right)\right) \\
-\frac{1}{2}p_{ct}\delta_0^3
\frac{s_-^2}{(E(CT)-E(P^*(-)))(E(P^{**}-E(P^*(-)))^2}
\left(1-c_+^2\frac{E(CT)-E(P^{**}}{E(CT)-E(P^*(-))}\right)^2 + \dots
\\
\alpha(\delta_0)=-p_{ct}^2 s_-^2 
\left(\frac{1}{E(CT)-E(P^*(-))}+s_+^2 \left(\frac{1}{E(P^{**})-E(P^*(-))}
-\frac{1}{E(CT)-E(P^*(-))}\right)\right) + \\
-\frac{3}{2}\delta_0^2p_{ct}^2\frac{s_-^2}
{(E(CT)-E(P^*(-)))(E(P^{**}-E(P^*(-)))^2}
\left(1-c_+^2\frac{E(CT)-E(P^{**}}{E(CT)-E(P^*(-))}\right)^2 + \dots

\end{array}
\end{equation}

which shows the effective permanent dipole moment and polarizability of the lower dimer band 
For a symmetric dimer the dipole moment  is zero and the polarizability is approximately given by

\begin{equation}
\alpha_0=-s_-^2 \frac{p_{ct}^2}{E(CT)-E(P^*(-))}
\end{equation}

The dipole moment as well as the polarizability grow with increasing asymmetry. 
The higher order terms  are small as long as the asymmetry $\delta$ is small compared to
 the energy difference $E(P^{**}) - E(P^*(-))$. For the special pair  we estimate
 $E(P^{**}) - E(P^*(-)) = 3000 cm^{-1}$. The asymmetry which is necessary to explain the experimental
 Stark effect [9] is of the same order. Hence the higher order terms are in fact important.
The Stark effect is strongly enhanced via the interaction with the state $P^{**}$. For maximum
 mixture of the $P^{**}$ and CT(+) states we have $s_+^2 =1/2$.  For 
$E(CT)-E(P^*) = 6000 cm^{-1} $ and
$ E(P^{**})-E(P^*)=3000cm^{-1} $ this leads
 to an increase of $50\%$ both for the dipole change and 
the polarizability.  For $s_-^2 =0.14$  the calculated dipole change is without this enhancement
 effect $ (0.73  (\delta /1000 cm^{-1}) + 0.04  (\delta/1000 cm^{-1})^3) Debye$  
 which yields a value
 of $5.5 Debye$ for $\delta=4000cm^{-1} $.

In our INDO calculations such a large splitting resulted only
 for the acetyl rotated structure. Interaction with the $P^{**}$ state increases the 
permanent dipole moment by $17\%$ to $6.4 Debye$. This is roughly the correct magnitude
 consistent with our earlier assignment. The analysis of the experimental spectra, however, 
is complicated by its shape which contains contributions from the first and second derivative.
 Figure 1c shows a simulation using the calculated couplings from table 2 and the same 
adjusted energies as in figure 1b. The calculated Stark effect contains a significant 
contribution from the large polarizability of the dimer which shows up as a first derivative. 
We like to mention that our model also allows to calculate the higher order polarizabilities
 which have been investigated  in recent experiments [ 11]. For the absorption changes 
proportional to the fourth power of the field our calculations predict a mixture of up 
to the fourth derivative of the absorption spectrum with the major contribution from the
 third derivative. These seems to be in qualitative agreement with the experiment.

\subsection*{
Conclusion}
Our model calculations strongly suggest that a double-triplet excitation $P^{**}$ of the dimer 
is in the energy region of the internal charge transfer states and is strongly coupled to them.  
We identify this state with the narrow band of the experimental $P^*$ spectrum at $2700cm^{-1}$. This 
interpretation  explains the intensity and the relative small band width as compared to the broad
 band at higher energies which is typical for CT excitations. Furthermore  mixture of $P^{**}$ and the 
CT states enhances the Stark effect which helps to explain the experimental data. The relatively 
strong coupling involves the transition of an electron from an occupied orbital located on one
 half to an unoccupied orbital on the other half. This coupling is largely analogous to the 
coupling of the lowest cationic state of $P^+$ to a trip-doublet state which has been  invoked 
to explain the experimental $P^+$ spectrum. Our  INDO calculations are based on a structure model 
where the acetyl of $P_M$ is rotated by $180^o$ . This way the proper splitting of the CT states could be obtained.

\subsection*{Acknowledgement}
This work has been supported by the Deutsche Forschungsgemeinschaft (SFB 143 and SFB 533) .

\subsection*{
References}

\begin{enumerate}
\item
	K.Wynne, G.Haran, G.D.Reid, C.C.Moser, P.L.Dutton, R.M.Hochstrasser, 
	J.Phys.Chem. 100(1996) 5140-5148
\item
	P.O.J.Scherer, S.F.Fischer, in: The Photosynthetic Bacterial 
Reaction Center Center II, eds. J.Breton 	
and  A.Verméglio, Plenum Press New York 1992, p 193-207 

\item
	J.Breton, E.Nabedryk, W.W.Parson , Biochemistry 31 (1992) 7503-7510

\item
	W.W.Parson, E.Nabedryk, J.Breton, in:The Photosynthetic Bacterial Reaction Center II, eds. J.Breton 
	and A.Verméglio, Plenum Press New York 1992, p.79-88

\item
	J.R.Reimers, N.S.Hush, J.Am.Chem.Soc. 117 (1995) 1302-1308

\item
	M.Plato, W.Lubitz, F.Lendzian, K.Möbius, Isr.J.Chem. 28 (1988) 109

\item
	J.M.Ohlson, J.Trunk, J.L.Sutherland,, Biochemistry 24 (1985) 4495

\item
	P.O.J.Scherer, S.F.Fischer, in : perspectives in photosynthesis, 
ed. J.Jortner and B. Pullman, Kluwer 	academic publishers 1990, 361-370

\item
	P.O.J.Scherer, S.F.Fischer, in: The Reaction Center of 
Photosynthetic Bacteria III, 
	ed. M.E.Michel-Beyerle, Springer New York 1995, p.89-104

\item
	P.O.J.Scherer, C.Scharnagl, S.F.Fischer, Chem.Phys.Lett. 197 
(1995)333-341

\item
	K.Lao, L.J.Moore, H.Zhou, S.G.Boxer, J.Phys.Chem. 99 (1995) 496-500

\item
	B.S.Hudson, B.E.Kohler, Chem.Phys.Lett. 14 (1972) 299

\end{enumerate}

\subsection*{
Figure caption}

figure 1: simulated spectra from  the exciton model using the parameters from table 3 

(a) the experimental $P^+$ spectrum  from [3] is compared with the calculated transitions shown as bars. As compared to the INDO results the HOMO-HOMO coupling has been enlarged by a factor of 1.8 to reproduce the position of the hole transfer band and the doub-triplets are shifted by  $2800 cm^{-1}$.

(b) the calculated  $P^*$ spectrum (full curve) is compared with  a fit of the experimental data by two Gaussians (broken curve) as described in [1]. The scale is the same for both curves. The higher intensity seen experimentally for  the lower band could be explained if the coupling of the CT states to $P^{**}$ and $P^*$(-) were enlarged as compared to the calculated values.

(c) the calculated Stark effect spectrum (curve) is shown together with the calculated absorption (reduced by a factor 0.05). The dimer band shows a mixture of first and second derivative with the minimum close to the position of the absorption maximum. The other excited states contribute only little to the Stark effect. The width of the dimer bands are as in earlier  simulations of the room temperature spectra. The splitting of the CT states of $4000 cm^{-1}$ is from INDO results on the M-acetyl rotated structure. The coupling of $P^{**}$ and the CT states is essentially the same as the coupling  between the trip-doublets and the lowest cationic states of  the dimer. The differences in table 3 are due to different reorganization effects.

.

\begin{table}[p]
\begin{center}
\begin{tabular}{c|c}
zero order & with CI \\
\hline
$U_1=-\beta_{LM}=320cm^{-1}$ & $U_1=800cm^{-1}$ \\
$U_2=\frac{1}{2}\beta_{L*M*}=-970cm^{-1}$ & $U_2=-970cm^{-1}$ \\
$U_{12}=-\sqrt{\frac{3}{2}}\beta_{L*M}=565 cm^{-1} $ & $U_{12}=1370cm^{-1}$\\
$U_{21}=-\sqrt{\frac{3}{2}}\beta_{LM*}=1050 cm^{-1} $ & $U_{21}=1770cm^{-1}$\\
\hline
\end{tabular}
\caption{\it calculated coupling matrix elements of the lowest $P^+$ excitations. The left column shows the CI matrixelements of the main configurations as a function of one electron resonance integrals $\beta$ between localized orbitals of the neutral dimer groundstate. L(M) and L*(M*) denote the highest occupied and
the lowest unoccupied orbitals of $P_{L(M)}$. The right column shows the
effect of the reorganisation of the electron system in the cationic state.
}
\end{center}
\end{table}

\begin{table}[p]
\begin{center}
\begin{tabular}{c|c}
zero order & with CI \\
\hline
$U_-=\beta_{LM}+\beta_{L*M*}=-2180cm^{-1}$ & $U_-=-2100cm^{-1}$ \\
$U_+=\beta_{L*M*}-\beta_{LM}=-1530cm^{-1}$ & $U_+=-202cm^{-1}$ \\
$U_d=-\sqrt{\frac{3}{2}}(\beta_{LM*}+\beta_{L*M}=1690cm^{-1}$ & 
$U_d=1370cm^{-1}$ \\
\hline
\end{tabular}
\caption{\it calculated coupling matrix elements for the lowest excited
singlet states of the dimer. The left column shows the one electron
resonance integrals of the main configurations. The right column
shows the dynamic correlation effect.}
\end{center}

\end{table}

{\footnotesize
\begin{table}[p]
\caption{\it parameters of the simulation fig. 1}
\begin{tabular}[t]{cccc}
\hline
$L^+$ & $M^+$ & $L^+M^3$ & $M^+L^3$ \\
\hline
0 & -1290 & 0 &  -1370 \\
 & 1610 & -1770 & 0 \\
 &  &  6690 & 970 \\
 & & & 7500\\
\hline
\multicolumn{4}{c}{coupling matrix for$ P^+$ in $cm^{-1}$}
\end{tabular}
\begin{tabular}[t]{ccccc}
\hline
$P^*(-)$ & $P^*(+)$ & $L^+M^-$ & $M^+L^- $ & $P^{**}$ \\
\hline
0 & 0 & 1370 & -1610 & 0 \\
 & 480 & 320 & -80 & 0 \\
& & 2500 & 0 & 800 \\
& & & 6530 & 1130 \\
& & & & 2500 \\
\hline
\multicolumn{5}{c}{coupling matrix for$ P^*$ in $cm^{-1}$}
\end{tabular}
\begin{tabular}[t]{cc}
\hline
$P^*(-)$ & 370 \\
$P^*(+)$  &  180 \\
$P^{**}$ &  3000 \\
\hline
\multicolumn{2}{c}{linewidth FWHM in $cm^{-1}$}
\end{tabular}

\end{table}
}
\normalsize

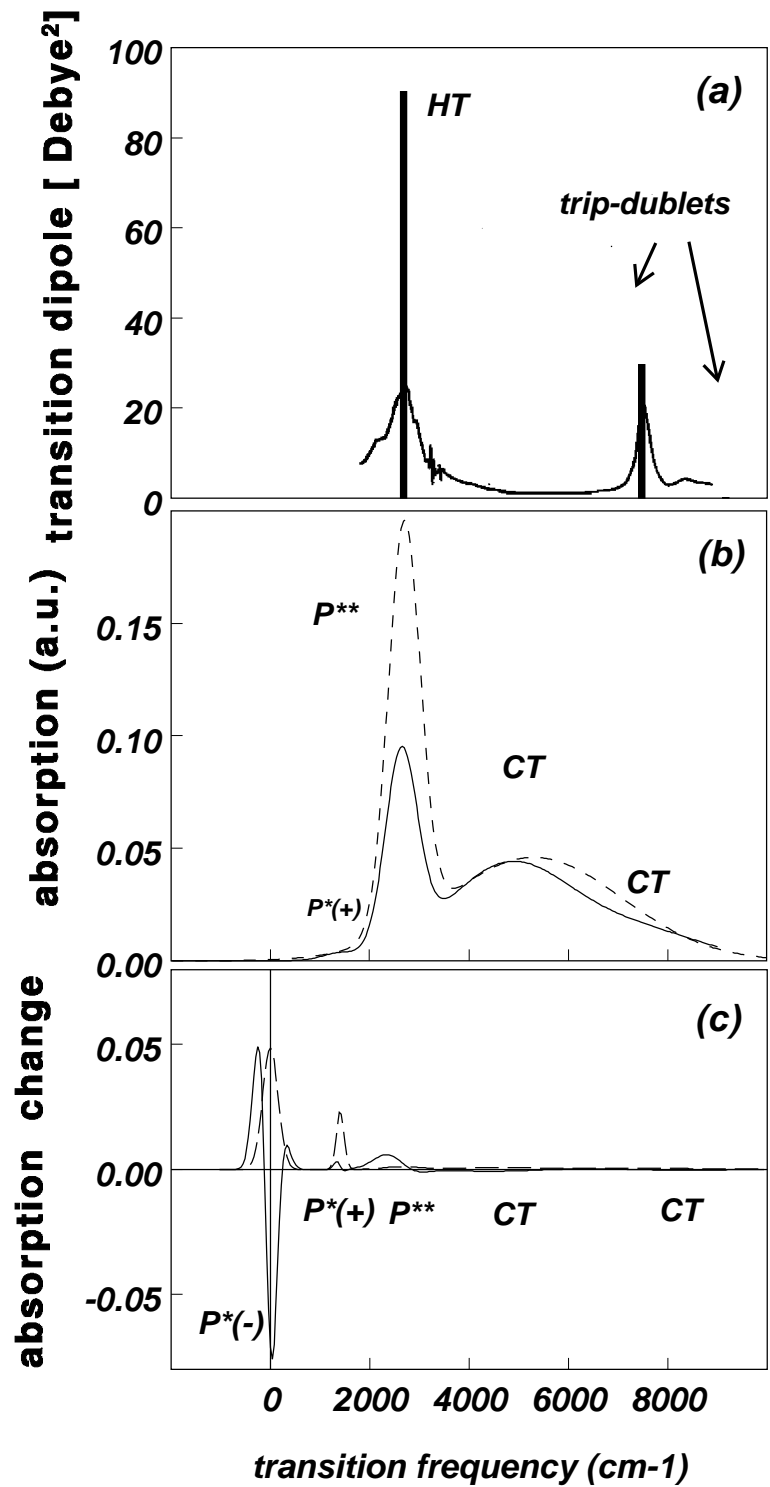
\end{document}